\documentclass[12pt,letterpaper]{article} 

\usepackage[margin=0.8in,includefoot=false,
            footskip=0.4in,nohead=true]{geometry} 

\usepackage{url}
\usepackage{hyperref}
\usepackage[dvipsnames]{xcolor}
\usepackage{algpseudocode}
\usepackage{algorithm}
\usepackage[numbers,sort&compress]{natbib} 
\usepackage{verbatim}
\usepackage{authblk}
\usepackage[frak=mma,scr=rsfs,cal=mma]{mathalfa}
\DeclareMathAlphabet{\mathcal}{OMS}{cmsy}{b}{n}
\usepackage[pdftex]{graphicx}
\usepackage{wrapfig}
\usepackage{enumitem}
\usepackage{tabularx}
\usepackage[margin=6pt,font={sf,scriptsize},labelfont=bf]{caption}
\usepackage{times}
\usepackage{amsmath}
\usepackage{amssymb}
\usepackage{amsfonts} 
\usepackage{dsfont}
\usepackage{float}
\definecolor{linkcolour}{rgb}{0,0.2,0.6}
\definecolor{linkcolour2}{rgb}{0.7,0.1,0.1}
\hypersetup{colorlinks,breaklinks,linkcolor=linkcolour2,citecolor=linkcolour,filecolor=linkcolour,urlcolor=linkcolour}
\usepackage{sectsty}
\usepackage{multirow}
\usepackage{breakcites}
\usepackage{booktabs}
\usepackage{caption}
\usepackage{longtable}
\usepackage[flushleft]{threeparttable}
\usepackage{lipsum}

\newcommand\blfootnote[1]{%
  \begingroup
  \renewcommand\thefootnote{}\footnote{#1}%
  \addtocounter{footnote}{-1}%
  \endgroup
}

\sectionfont{\large \sffamily\bfseries}
\subsectionfont{\normalsize \sffamily\bfseries}
\subsubsectionfont{\it \sffamily}
\setlist[itemize]{noitemsep, topsep=2pt, labelsep=3mm} 
\setlist[enumerate]{noitemsep, topsep=2pt, labelsep=3mm} 

\let\olditemize =\itemize
\let\oldenumerate =\enumerate
\let\olddescription =\description
\def\Nospacing{\itemsep=1pt\topsep=0pt\partopsep=0pt%
\parskip=0pt\parsep=0pt}
\def\noitemsep{
\renewenvironment{itemize}{\olditemize\Nospacing}{\endlist}
\renewenvironment{enumerate}{\oldenumerate\Nospacing}{\endlist}
\renewenvironment{description}{\olddescription\Nospacing}{\endlist}
}
\makeatletter

\usepackage{titlesec}
\titlespacing*{\section}{0pt}{0.8\baselineskip}{0.5\baselineskip}
\titlespacing*{\subsection}{0pt}{0.5\baselineskip}{0.3\baselineskip}
\titlespacing*{\subsubsection}{0pt}{0.3\baselineskip}{0.2\baselineskip}

\begin{document}

\title{\vspace{-30pt}
\Large\sffamily\bfseries UNADON: Transformer-based model to predict genome-wide chromosome spatial position}

\normalsize
\author[1]{\normalsize \sffamily Muyu Yang}
\author[1,*]{\normalsize \sffamily Jian Ma}
\affil[1]{\small \sffamily Computational Biology Department, School of Computer Science, \authorcr \small \sffamily Carnegie Mellon University, Pittsburgh, PA 15213, USA}
\affil[*]{\small \sffamily Correspondence:
\href{mailto:jianma@cs.cmu.edu}{jianma@cs.cmu.edu}}

\noitemsep
\date{}
\maketitle

\begin{abstract}
\noindent 
The spatial positioning of chromosomes relative to functional nuclear bodies is intertwined with genome functions such as transcription.
However, the sequence patterns and epigenomic features that collectively influence chromatin spatial positioning in a genome-wide manner are not well understood.
Here, we develop a new transformer-based deep learning model called UNADON, which predicts the genome-wide cytological distance to a specific type of nuclear body, as measured by TSA-seq, using both sequence features and epigenomic signals.
Evaluations of UNADON in four cell lines (K562, H1, HFFc6, HCT116) show high accuracy in predicting chromatin spatial positioning to nuclear bodies when trained on a single cell line. 
UNADON also performed well in an unseen cell type. 
Importantly, we reveal potential sequence and epigenomic factors that affect large-scale chromatin compartmentalization to nuclear bodies. 
Together, UNADON provides new insights into the principles between sequence features and large-scale chromatin spatial localization, which has important implications for understanding nuclear structure and function. 
\end{abstract}

\vspace{10pt}

\blfootnote{This article has been published in the ISMB 2023 Proceedings (\textit{Bioinformatics}, 39(Supplement\_1):553-i562), Oxford University Press. All rights reserved.}


\section*{Introduction}

The cell nucleus of higher eukaryotes is a heterogeneous organelle that contains distinct subnuclear structures and nuclear bodies associated with crucial cellular functions~\citep{spector2001nuclear,dundr2010biogenesis}.
For example, nuclear speckles regulate the transcription and splicing of genes~\citep{spector2001nuclear,chen2019genome} 
and the nuclear lamina anchors the chromatin towards the nuclear periphery~\citep{van2017lamina}.
Chromosomes are folded and packaged inside the nucleus, and they interact with various types of nuclear bodies, which has important implications for the structure and function of the higher-order genome organization. 
In particular, gene expression level is found to be inversely correlated with distance to nuclear speckles~\citep{zhang2021tsa}, and nuclear lamina frequently interacts with repressive chromatin~\citep{van2017lamina}. 
In the past decade, the rapid development of genome-wide mapping of chromatin interactions such as Hi-C~\citep{lieberman2009comprehensive} has revealed that the nuclear genome has multiscale structures, including A/B compartments~\citep{lieberman2009comprehensive} and subcompartments~\citep{rao20143d}, topologically associating domains (TADs)~\citep{dixon2012topological}, and chromatin loops~\citep{jin2013high, rao20143d}.
However, the interactions between large-scale chromatin and functional nuclear bodies in the nucleus remain poorly understood.
The recently developed tyramide signal amplification sequencing (TSA-seq) genomic mapping method~\citep{chen2018mapping} measures the cytological distances from chromosome loci to nuclear bodies, providing unprecedented opportunities to probe the spatial positioning of chromosomes relative to specific nuclear bodies in a genome-wide manner. 
Recent work has integrated different 3D genome mapping data to generate models of genome organization, including their distances to nuclear landmarks~\citep{boninsegna2022integrative}.
To reach a more complete understanding of the interplay between nuclear structure and function, it is imperative to reveal the underlying principles and determinants (e.g., sequence features and various local epigenomic signals) for chromatin-nuclear body interactions in a wide range of cellular contexts.
Unfortunately, there are currently no computational methods available to facilitate such analysis by combining genome sequence features and other genome-wide epigenomic features.

In the past few years, deep learning models have demonstrated strong prediction performance for functional genomic and epigenomic features~\citep{eraslan2019deep,yang2022machine}.
Specifically, deep learning architectures have been applied to predict 3D genome features in different scales based on genomic sequence features~\citep{yang2022machine}.
Computational methods such as Akita~\citep{fudenberg2020predicting} integrate DNA sequences up to 1Mb by using a series of dilated convolutional neural network (CNN) layers to expand the receptive field.
However, this approach may miss important dependencies for spatial positioning domains that can be as large as 10Mb~\citep{briand2020lamina}.
A more recent method Orca~\citep{zhou2022sequence} expands the maximum input sequence length to the whole chromosome by employing a large CNN with a cascading prediction mechanism.
However, the ability of CNN-based models to capture long-range dependencies is restricted by the size of the receptive field.
Deeper CNNs with larger dilation rates are required for large-scale chromatin feature predictions with long sequences, which is computationally inefficient to train.
Besides, very few existing models combine both genomic sequence and epigenomic features to effectively perform cross-cell-type predictions. 

Here, we introduce a new multi-modal transformer-based deep learning model called UNADON, which predicts chromatin spatial positioning relative to nuclear bodies based on DNA sequences and epigenomic signals.
The major contributions of our work are as follows:
(1) UNADON is a deep learning-based model to specifically predict chromatin spatial positioning relative to nuclear bodies;
(2) The distinctive neural architecture design enables UNADON to learn the long-range dependencies more effectively;
(3) UNADON generalizes well in the cross-cell-type predictions, which can be applied to infer spatial positioning in new cell types.
(4) Interpretation of UNADON reveals potential mechanisms for targeting nuclear bodies.
Together, UNADON establishes a novel computational framework for the prediction and interpretation of the spatial positioning of chromosomes.


\section*{Methods}

\subsection*{Overview of UNADON}

\begin{figure}[!htb]
\centering
\includegraphics[width=0.86\textwidth]{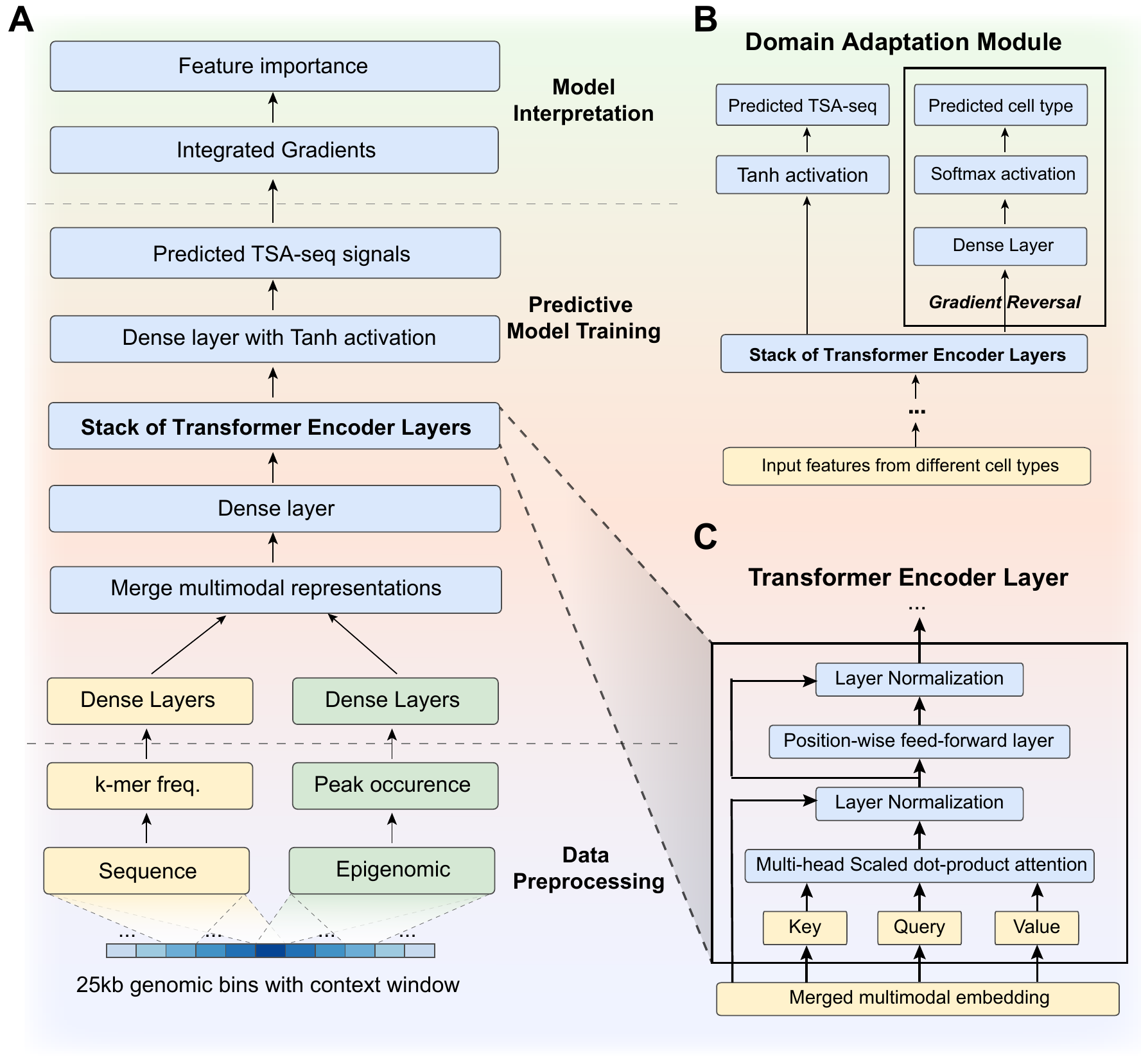}
\caption{
\textbf{(A)} Overview of UNADON. 
The workflow consists of three steps. 
(1) Data preprocessing. For each genomic region of 25kb, the DNA sequence feature is represented by a vector of k-mer frequency. 
The epigenomic feature is represented by the occurrence of epigenomic signal peaks within the region. 
(2) Predictive model training. 
To effectively incorporate different data modalities and large-context information, the model adopts a multi-modal design and employs the transformer modules with self-attention mechanisms.
(3) Model interpretation. We evaluate the important sequence elements and epigenomic signals using the feature contribution score computed by Integrated Gradients.
\textbf{(B)} An illustration of the domain adaptation techniques applied on top of UNADON in the cross-cell-type predictions. 
\textbf{(C)} The details of the transformer encoder layers.
}
\label{fig:1}
\end{figure}

UNADON is a transformer-based deep learning framework for predicting the spatial positioning of chromosomes relative to nuclear bodies from DNA sequence features and epigenomic signals. 
To integrate the input data of different modalities and enhance model interpretability, we highlight the following three distinctive features of UNADON: 
(1) Preprocessing of the sequence and epigenomic signals enables efficient learning of complex features.
(2) The multi-modal design allows the network to extract features for sequence and epigenomic signals separately; 
(3) The transformer modules capture crucial long-range dependencies of genomic bins.

The workflow of UNADON comprises three components: data preprocessing, predictive model training, and model interpretation (see \textbf{Fig.}~\ref{fig:1}A for an overview).
The genome is segmented into non-overlapping bins as the basic units for prediction and interpretation.
The model considers all genomic bins within a certain window size as input to provide context information.
The processed DNA and epigenomic features are fed into two subnetworks consisting of dense layers to learn the sequence and epigenomic representations independently, which are merged and passed into the transformer encoder modules.
The most important component of the transformer encoder modules is the multi-head attention layers.
The attention layers automatically learn the relevance of the neighboring bins in predicting the TSA-seq signal, allowing the model to incorporate long-range information.
After training and evaluation, a post-hoc feature attribution method, Integrated Gradients, is applied to systematically estimate the importance of loci and the contribution of different features.
The source code of UNADON can be found at: \url{https://github.com/ma-compbio/UNADON}.

\subsection*{Neural network architecture in UNADON}

The network architecture for UNADON is shown in \textbf{Fig.}~\ref{fig:1}A. 
The details of the architecture are as follows. 
For predicting the TSA-seq signal for one 25kb genomic bin, we consider the large context of +/- 2.5Mb.
The pre-processed sequence and epigenomic features from the context window are passed to two separate feature extraction sub-networks composed of dense layers with ReLU activation.
The sequence and epigenomic feature embeddings are then merged by concatenation and fed into the transformer encoder modules, which incorporate the long-range context information by learning the pairwise dependencies between all the input loci. 

The transformer encoder module is adapted from~\citep{vaswani2017attention}.
The architecture is described in \textbf{Fig.}~\ref{fig:1}C, 
which consists of one multi-head attention layer and one position-wide feed-forward layer. 
The multi-head attention layer trains multiple attention heads independently to attend to different parts of the sequence. 
\begin{equation}
\text{Multihead}(Q,K,V) = \text{Concat}(\text{head}_1,..., \text{head}_h)W^O 
\end{equation}
\begin{equation}
\text{where } \text{head}_i = \text{Attention}(QW^Q_i,KW^K_i,VW^V_i)
\end{equation}
Note that $W^Q$, $W^K$, $W^V$, and $W^O$ are projection matrices with trainable parameters.
The attention mechanism uses the scaled dot product of the key and query to calculate the relevance between two loci.
Specifically, it is defined as:
\begin{equation}
\text{Attention}(Q,K,V) = \text{softmax}\left (\frac{QK^T}{\sqrt{d_k}}\right )V
\end{equation}
The raw attention weight $QK^T$ is normalized by the factor $\sqrt{d_k}$, which is the dimension of the keys, and a softmax function.
Additionally, layer normalization is applied after each layer for better generalization.
Residual connections allow the gradient to flow directly through the layers to mitigate the vanishing gradients problem. 
A dense layer with Tanh activation function is applied to the output embedding of the transformer to predict the TSA-seq signals. 
To incorporate the positional information of the genomic bins, we adapted the relative positional embedding method introduced in TransformerXL \citep{dai2019transformer, labml}.

For cross-cell-type predictions, we attached a domain adaptation module~\citep{ganin2015unsupervised} to prevent the model from overfitting on the training set (\textbf{Fig.}~\ref{fig:1}B).
This domain adaptation module is an additional classifier that identifies the source cell types of the embeddings.
The forward pass leverages the cell-type-specific information within the embeddings to predict the source cell type.
The backward pass contains a gradient reversal layer that reverses the gradient flowing back into the transformer.
As a result, the model will maximize the loss for cell type prediction, thus discouraging the learning of any cell-type-specific features that can be used to distinguish between the cell types.

\subsection*{Model training and evaluation}

For all experiments, we adopt a cross-chromosome evaluation approach, where odd-numbered chromosomes are used for training and validation, and even-numbered chromosomes are held out for testing.
To search for the optimal hyperparameters, we conducted cross-validation by holding out one odd-numbered chromosome as a validation set for each fold.
The models were trained by minimizing the mean squared error (MSE) for the prediction of TSA-seq signals using AdamW optimizer \citep{loshchilov2017decoupled}. 
Learning rate warmup and decay were defined in a similar way as described in~\cite{vaswani2017attention}.
We recorded the validation loss for each epoch and used the model parameter which yields the minimum validation metrics as the final model for downstream evaluation on the testing sets and model interpretation.

To evaluate the performance and the generalizability of the model, we ran two types of evaluations: single-cell-type prediction and cross-cell-type prediction on four human cell lines (K562, H1, HCT116, HFFc6). 
For single-cell-type prediction, the model is trained on the sequence and epigenomic features in one single cell type. 
For cross-cell-type prediction, one cell type is held out from the training to estimate the performance of our model on an unseen cell type in the real world.
The model is trained on the odd-numbered chromosomes of three cell types and tested on the even-numbered chromosomes of the held-out cell type.

As a performance comparison, we also trained and evaluated the following four machine learning models. 
XGBoost is an optimized library for gradient tree boosting (GTB), which is a simple but powerful decision tree-based machine learning model~\citep{chen2016xgboost}. 
Dense neural networks (DNN) consist of fully connected layers that perform nonlinear transformations to the input. 
XGBoost and DNN predict the signal without considering any context information.
CNNs contain convolutional layers that are able to incorporate local information from neighboring bins. 
They have been extensively used in sequence-based predictive models, such as DeepSEA~\citep{zhou2015predicting}. 
However, different from the previous methods that predict the signal from raw DNA sequence, CNN is applied on the pre-engineered sequence features instead of the one-hot encoding of sequences. 
In addition, we included dilated CNNs, which boost the performance of CNNs by increasing the size of the context that can be incorporated.
To ensure a fair comparison, we have utilized the same input features and context length for the CNN and dilated CNN models as we have for UNADON.
The machine learning algorithms were implemented using the scikit-learn library~\citep{scikit-learn}.
The neural network-based models were implemented using PyTorch~\citep{paszke2019pytorch}.

For cross-cell-type predictions, we evaluated an additional baseline model inspired by~\cite{schreiber2020pitfall}.
Considering that the spatial positioning of chromosomes shares some similarities across cell types, it is reasonable to use the average signals of existing cell types as an estimation for an unseen cell type.

\subsection*{Interpretation of neural network} \label{sec:interp}

To reveal the contribution of the input features to the final predictions, we applied Integrated Gradients~\citep{sundararajan2017axiomatic, kokhlikyan2020captum} to derive the locus-specific importance scores for sequence and epigenomic signals separately.
Integrated Gradients is a gradient-based feature attribution method that calculates the importance scores based on the gradient of the output compared to the input features.
Intuitively, if a small deviation from the baseline causes a big change in the output, the feature should be important for the prediction.
Integrated Gradients considers a straight-line path between the input and the baseline.
The importance score is defined by accumulating the gradient of all points on this path.
We used the vector of all zeros as the baseline.
Since importance scores computed by Integrated Gradients can be either positive or negative based on the effect of the features on the predictions, we took the absolute value of the importance scores as feature contribution.
To quantify the total contribution of DNA sequence features, we aggregated the contribution from all dimensions by summing up the importance scores.

\subsection*{Data collection and preprocessing} \label{sec:preprocessing}

We collected the genome-wide mapping of cytological distance to nuclear bodies from TSA-seq data in four human cell lines (K562, H1, HCT116, HFFc6)~\citep{dekker20174d}. 
We followed the same normalization process as discussed in ~\citep{zhang2021tsa}, which converted the normalized read counts to the TSA-seq enrichment scores.
We calculated the TSA-seq scores for non-overlapping 25kb genomic bins and scaled the signal between -1 and 1. 
To alleviate the effect of technical noise, we smoothed the TSA-seq signal using a Hanning window of size 21.
Higher signals represent a closer distance to a specific type of nuclear body.

We used the human reference genome GRCh38. 
Genomic regions with low mappability, such as the centromere region, were removed.
To reduce the dimensionality of DNA sequences, we use k-mer frequency to represent the sequences.
We merged k-mers with their reverse complements to consider both the forward and the reverse strands of the DNA.
We concatenated the k-mer frequencies for $k = 5$ and $k = 6$ and applied PCA to reduce the dimension of the k-mer frequency vector to 20.

To capture the cell-type-specific chromatin spatial positioning, we incorporated the widely available epigenomic features. 
Specifically, we aggregated chromatin accessibility and eight histone modifications (H2A.Z, H3K4me1, H3K4me2, H3K4me3, H3K9me3, H3K27me3, H3K27ac, H3K36me3) from different assays and database~\citep{encode2012integrated,dekker20174d,janssens2018automated}, as summarized in Supplementary Table.
We used the peaks of the signal called by the ENCODE peak calling pipeline~\citep{encode2012integrated} to represent the epigenomic signals within a 25kb genomic region.
For each individual region, we count the number of bases underlying the signal peaks and normalize it by the length of the region.
To account for potential technology-related bias, we further normalize the peak occurrence by the overall average peak frequency on the whole genome.


\begin{figure}[!htb]
\centering
\includegraphics[width=0.98\textwidth]{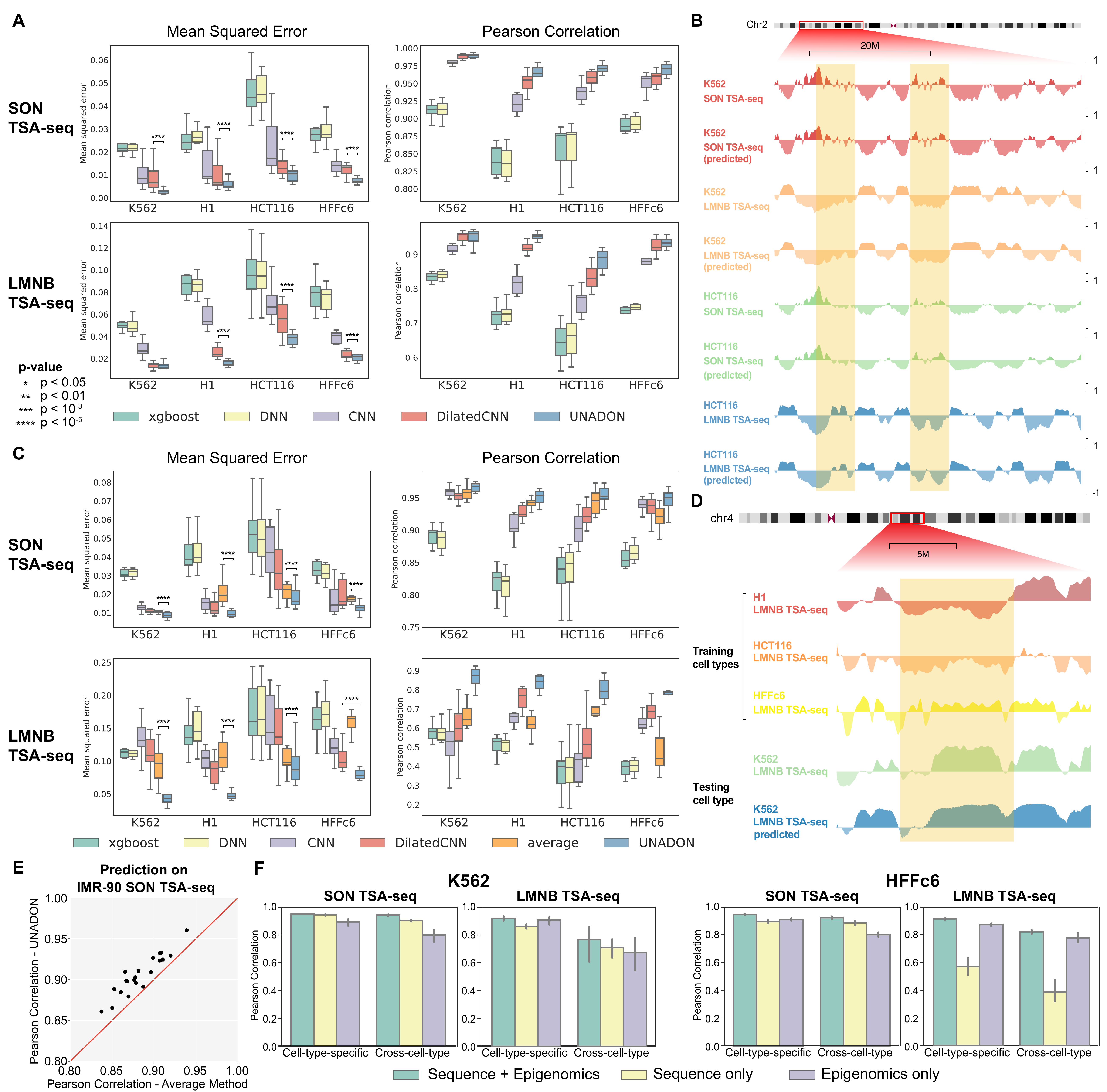}
\caption{Performance evaluation and comparison for the prediction of SON and LMNB TSA-seq signal using different baseline methods in K562, H1, HCT116, and HFFc6.
The boxplots show the mean squared error and Pearson correlation evaluated across the even-numbered chromosomes, which are held out for testing.
\textbf{(A)} Performance of single-cell-type models measured by mean squared error and Pearson correlation in individual cell types. 
The p-values for performance improvement are calculated based on the squared error per genomic locus.
\textbf{(B)} An illustration of actual and predicted TSA-seq in K562 and HCT116. 
UNADON is able to make accurate predictions for both SON TSA-seq and LMNB TSA-seq, as shown in the highlighted regions.
\textbf{(C)} Performance of cross-cell-type models measured by mean squared error and Pearson correlation. 
The models were trained on the odd-numbered chromosomes of three cell types and tested on the even-numbered chromosomes of one cell type.
The x-axis labels are the testing cell types.
\textbf{(D)} An illustration of cross-cell-type prediction results. 
The model accurately predicts the highlighted cell-type-specific patterns of K562 LMNB TSA-seq that do not appear in any of the training cell types.
\textbf{(E)} Prediction performance on external IMR-90 SON TSA-seq. 
The scatterplot shows the performance comparison of the average method and UNADON measured by Pearson correlation.
The dots represent the 22 autosomes in IMR-90.
\textbf{(F)} Ablation studies to demonstrate the importance of both sequence and epigenomic features for the prediction.
}
\label{fig:2}
\end{figure}

\section*{Results}


\subsection*{UNADON accurately predicts chromatin spatial positioning relative to nuclear bodies in individual cell type}

We applied UNADON to predict the spatial positioning of chromosomes relative to nuclear speckle (measured by SON TSA-seq) and nuclear lamina (measured by LMNB TSA-seq) in four human cell lines (K562, H1, HCT116, and HFFc6) (\textbf{Fig.}~\ref{fig:2}A).
Examples of the comparison of the predicted signals and the actual TSA-seq tracks are shown in \textbf{Fig.}~\ref{fig:2}B.
The highlighted region illustrates that UNADON accurately predicts the cell-type-specific patterns of the spatial positioning of chromosomes relative to specific nuclear bodies.
Importantly, even the small structures, such as the tiny peaks and valleys smaller than 1Mb, are predicted correctly.

We compared the performance of UNADON with four commonly used baseline methods: XGBoost, DNNs, CNNs, and Dilated CNNs (see \textbf{Methods}) in terms of mean squared error (MSE) and Pearson correlation coefficient (PCC).
We found that UNADON consistently outperforms the baseline models in all cell types and nuclear bodies (\textbf{Fig.}~\ref{fig:2}A).
Specifically, UNADON reaches a median Pearson correlation of 0.97 on SON TSA-seq and 0.92 on LMNB TSA-seq across all four cell types, 
which is significantly higher than XGBoost and DNNs with no context information.
UNADON also outperforms the basic CNN in all prediction tasks, suggesting that long-range dependencies are important for the prediction. 
Remarkably, UNADON achieves better prediction performance than the state-of-the-art dilated CNN, further demonstrating the strength of the attention mechanism in capturing the relationship between genomic features and the spatial positioning.
Note that the prediction performance of UNADON differs across the cell types, especially on LMNB TSA-seq.
The PCC in HCT116 LMNB TSA-seq is only 0.88, as compared with 0.95 in K562 LMNB TSA-seq. 
The discrepancy indicates that the cell types may contain distinct sequence and epigenomic features for large-scale chromosome positioning, which increases the difficulty of cross-cell-type prediction.

Overall, these evaluations confirm the advantage of UNADON over the baseline methods for predicting the spatial positioning of chromosomes in individual cell types.

\subsection*{UNADON can predict chromatin spatial positioning in unseen cell types}

To examine the ability of UNADON to generalize to unseen cell types, we conducted cross-cell-type evaluations by holding out one cell type for testing and training the models on the remaining three cell types.
We included an additional baseline method, named `average', which simply utilizes the average TSA-seq in training cell types as the prediction.
The average method also serves as a measure of how similar the spatial positioning is in the testing cell type compared with that in the training cell types.
Similar to the single-cell-type models, we evaluated the performance by calculating the mean squared error and Pearson correlation.
The advantage of UNADON over the baseline models becomes more pronounced in the cross-cell-type setting, as is shown in \textbf{Fig.}~\ref{fig:2}C.
SON TSA-seq is highly conserved among the cell types, as is revealed by the strong performance of the average method (median PCC 0.94) and the conserved example regions in \textbf{Fig.}~\ref{fig:2}B.
As a result, all baseline models reach a PCC above 0.8 on predicting cross-cell-type SON TSA-seq, where UNADON is the highest in all evaluations (median PCC 0.956).

LMNB TSA-seq is more variable across the cell types, as indicated by poor PCC (0.62) of the average method and the highly variable example regions in \textbf{Fig.}~\ref{fig:2}B.
Regardless, UNADON, with a median PCC between 0.785 (HFFc6) and 0.875 (K562), significantly outperforms all baseline methods, including CNNs and dilated CNNs that incorporate context information.
The result suggests that the transformer modules and the domain adaptation design are crucial for the generalizability of the models.
Additionally, we provided an example visualization of the K562 LMNB TSA-seq signals predicted by the cross-cell-type UNADON.
We also included the tracks from the training cell types as references.
Based on \textbf{Fig.}~\ref{fig:2}D, it is clear that UNADON is capable of predicting the distinct patterns of K562 LMNB TSA-seq, further confirming that UNADON is indeed learning the fundamental connections between the spatial positioning and the genomic features.

To further demonstrate the ability of UNADON to infer TSA-seq on a new cell type, we conducted an evaluation using an external IMR-90 SON TSA-seq dataset from a different lab~\citep{alexander2021p53}.
We reprocessed the raw data based on the TSA-seq preprocessing workflow (see \textbf{Methods}). 
UNADON achieves a mean Pearson correlation of 0.91 across all 22 chromosomes in IMR-90, which is consistently better than the performance of the average method, as is shown in \textbf{Fig.}~\ref{fig:2}E. 
The evaluation of UNADON on the external dataset provides additional validation of its ability to accurately predict TSA-seq in unseen cell types, supporting its practical utility and potential for inferring chromatin spatial positioning in new cell types in absence of TSA-seq data.

In addition, to probe the importance of different input features to the prediction performance, we conducted ablation studies by removing either sequence or epigenomic features from the input (\textbf{Fig.}~\ref{fig:2}F). 
For SON TSA-seq, sequence features alone are sufficient to reach a high prediction performance, suggesting the crucial role of sequence features in determining chromatin positioning towards nuclear speckles.
The incorporation of epigenomic features slightly boosts the performance by providing additional information of the chromatin states.
For LMNB TSA-seq, the result varies depending on the cell type.
In K562, the sequence-only models attain a moderate prediction performance.
However, in HFFc6, sequence features are significantly less predictive of the LMNB TSA-seq.
In this case, epigenomic features are necessary for the accurate prediction of the spatial positioning of chromatin.
These results suggest that different cell types may utilize distinct underlying mechanisms for targeting the chromatin toward nuclear lamina.
Therefore, both sequence and epigenetic features play critical roles in achieving the high prediction performance of UNADON.

Taken together, the cross-cell-type evaluations demonstrate that UNADON can still perform well in predicting chromatin spatial positioning in unseen cell types.

\subsection*{UNADON uncovers the genome-wide contribution of sequence and epigenomic features}  \label{sec:dis}

\begin{figure}[!htb]
\centering
\includegraphics[width=0.95\textwidth]{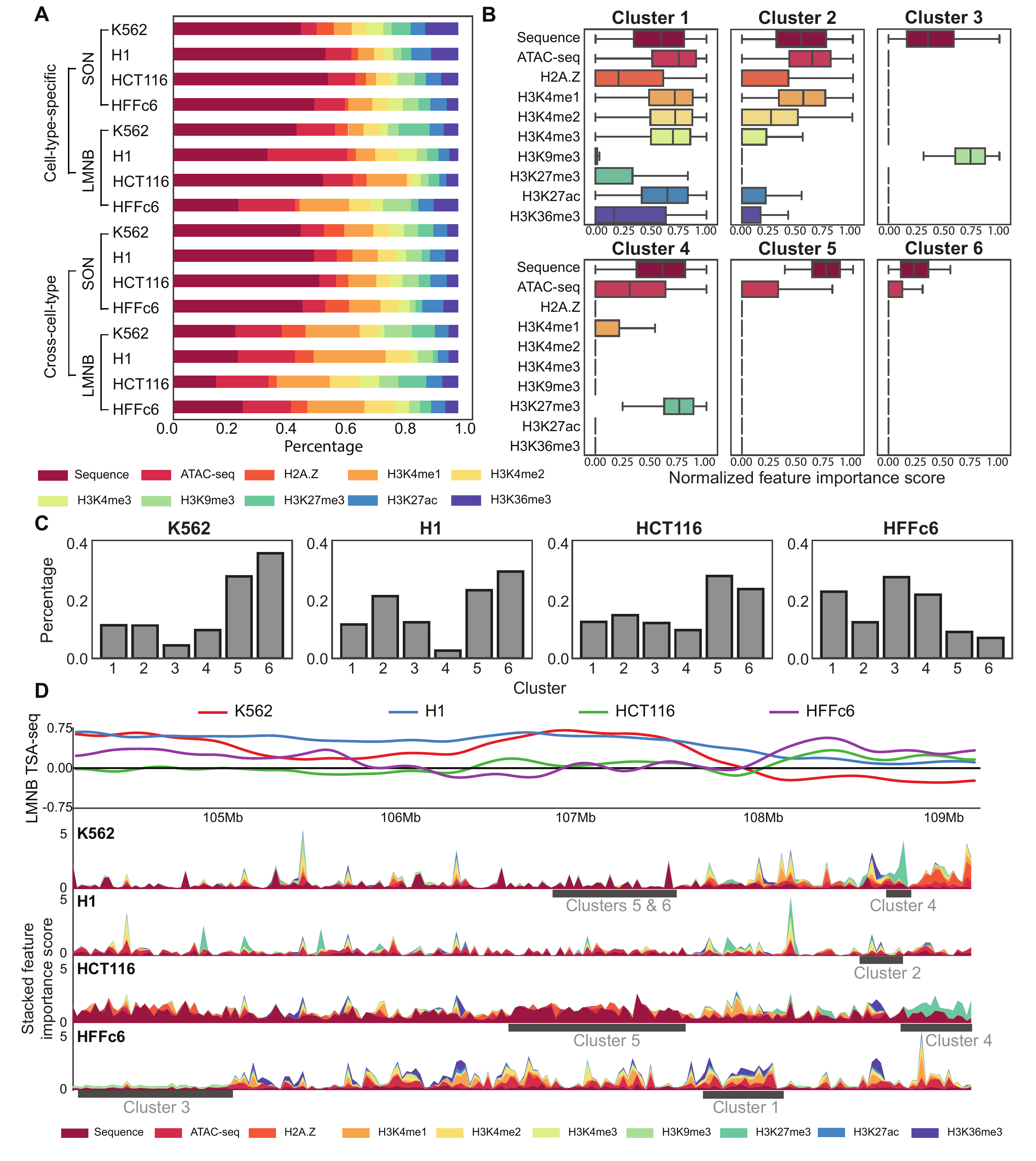}
\caption{The feature importance scores computed by Integrated Gradients delineate the genome-wide contribution of DNA sequences and epigenomic signals from LMNB TSA-seq across all four cell types.
\textbf{(A)} Percentage contribution of sequence and epigenomic features across all experimental setups. 
\textbf{(B)} Six distinct patterns of sequence and epigenomic contributions are identified by k-means clustering, each with a unique distribution of the normalized importance scores.
\textbf{(C)} Proportion of each pattern of feature contribution in the four cell types.
\textbf{(D)} Example of the distribution of feature importance score on chr4 across the four cell types.
The top track displays the LMNB TSA-seq signals.
The bottom tracks are stacked plots of feature importance scores.
The black bars underneath the important score tracks highlight the representative regions for each cluster of patterns.
}
\label{fig:3}
\end{figure}

We next applied Integrated Gradients to investigate the contribution of sequence and epigenomic features to the prediction of both SON TSA-seq and LMNB TSA-seq across the four cell types.
We first computed the mean importance score of each individual feature by averaging over all the genomic bins on the test chromosomes (\textbf{Fig.}~\ref{fig:3}A).
We found that overall the DNA sequence is the most important feature for prediction, followed by ATAC-seq and H3K4me1.
The distribution of feature importance varies drastically across different cell types and target nuclear bodies.
The contribution of sequence in predicting LMNB TSA-seq is comparatively less significant compared to its contribution to predicting SON TSA-seq, suggesting that the targeting mechanisms towards nuclear lamina may be less dependent on sequence features.
Another observation is that the contribution of epigenomic features in HFFc6 is greater than that in other cell types, which aligns with the findings from our ablation studies.
Moreover, the contribution of epigenomic features becomes more evident in the cross-cell-type settings, since the models need cell-type-specific epigenomic information for the predictions compared to the cell-type-agnostic sequence features.
This analysis shows a quantitative assessment of the role of sequence and epigenomic features in modulating chromatin spatial positioning toward nuclear bodies.

To gain insight into how different features collectively contribute to the prediction, we performed k-means clustering to identify patterns of the importance score distribution.
Given that the role of sequence and epigenomic features are more intertwined in the prediction of LMNB TSA-seq, we focus on the contribution scores derived from LMNB TSA-seq across four cell types.
The optimal number of clusters was determined through silhouette analysis, resulting in the identification of six distinct patterns (\textbf{Fig.}~\ref{fig:3}B).
The proportion taken up by each cluster is reported in \textbf{Fig.}~\ref{fig:3}C.
Cluster 1 has high sequence and epigenomic contributions from active histone marks, including H3K4me1, H3K4me2, H3k4me3, and H3K27ac.
Cluster 2 is characterized by the high contribution from ATAC-seq and H3K4me1, along with low to moderate contributions from H2A.Z and H3K4me2.
Cluster 3 has a unique high contribution from H3K9me3.
Cluster 4 is distinguished by a high contribution from H3K27me3.
Clusters 5 and 6 represent genomic regions where only DNA sequence is informative for the prediction.
Interestingly, cluster with low sequence and high epigenomic contribution is not detected, further emphasizing the essential role of sequence features.

Moreover, we show the distribution of feature importance scores in \textbf{Fig.}~\ref{fig:3}D in a 5Mb region to demonstrate how various cell types employ different patterns to predict cell-type-specific LMNB TSA-seq signals.
Representative regions for each cluster are annotated for each cluster at the bottom.
We observed that clusters 1 and 2 tend to be smaller, marked by the sharp peaks on the plot.
Clusters 5 and 6, with high and low sequence-only contributions respectively, are usually intertwined with each other, spanning a large region of the genome.

Overall, these results provide new insight into the underlying targeting mechanism of large-scale chromatin towards nuclear bodies, which will be further analyzed in the next section. 

\subsection*{UNADON identifies the potential sequence and epigenomic determinants for chromatin localization} \label{sec:exp}

\begin{figure}[!t]
\centering
\includegraphics[width=0.96\textwidth]{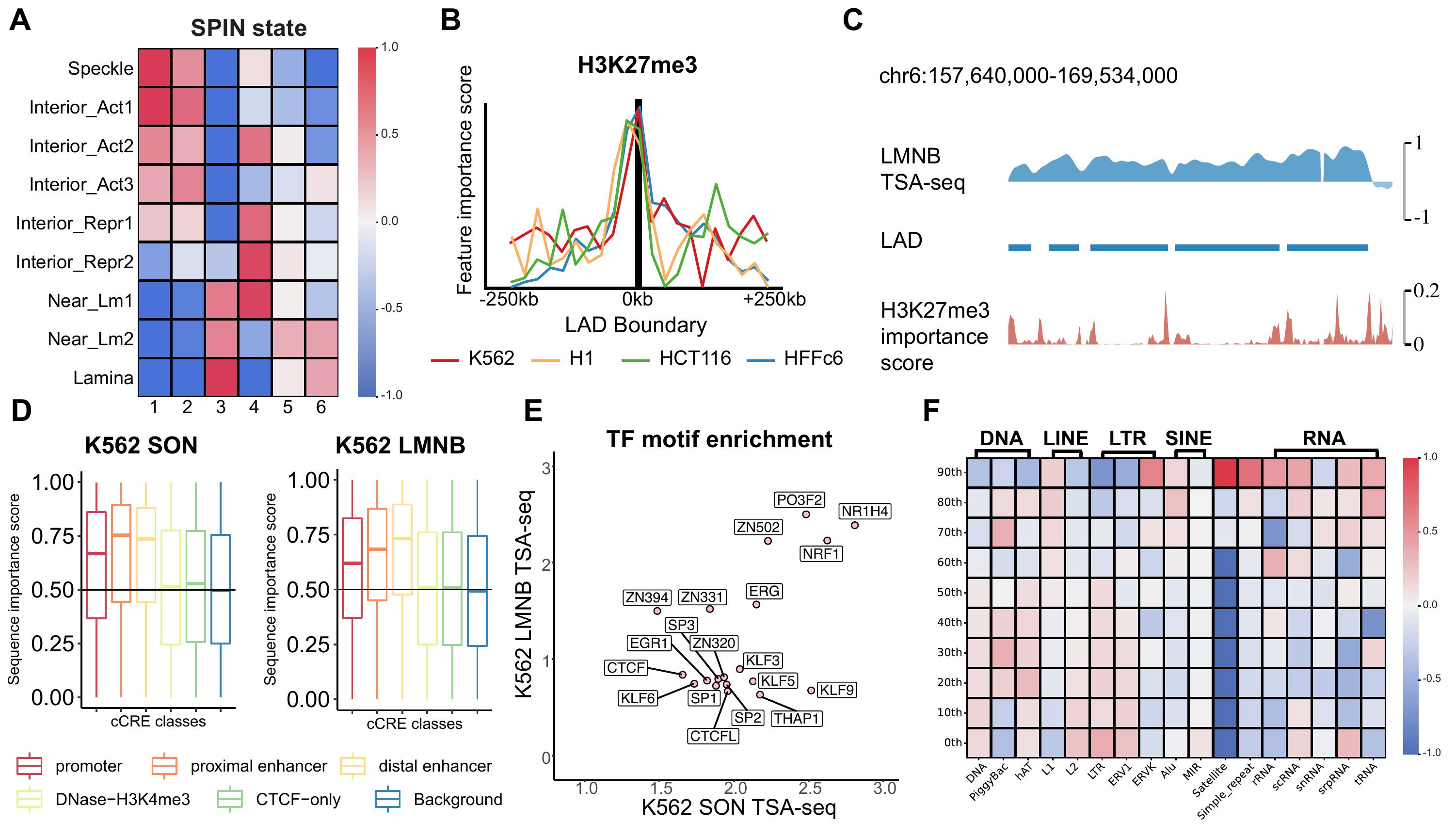}
\caption{
Analysis of the feature importance scores reveals the important sequence element and epigenomic features for chromatin spatial positioning relative to nuclear bodies.
\textbf{(A)} The comparison of the pattern clusters with SPIN states~\citep{wang2021spin}.
The colors represent the log2 fold change of the percentage of the SPIN states in a specific cluster relative to the randomly sampled background set.
\textbf{(B)} The distribution of H3K27me3 importance score near the LAD boundaries.
\textbf{(C)} Example of the correlation between LAD boundaries and H3K27me3 importance score on chr6: 157,640,000-169,534,000.
\textbf{(D)} The sequence importance scores in the top 5\% genomic regions with the highest enrichment of the cCREs annotated by SCREEN ~\citep{moore2020expanded}.
The randomly sampled background set is included as a reference.
\textbf{(E)} The TF motif enrichment log2 fold change of the top 10\% most important genomic regions for the prediction of K562 SON TSA-seq and K562 LMNB TSA-seq.
\textbf{(F)} The repetitive element enrichment fold change of genomic regions with different sequence importance levels.
The genomic regions in the test chromosomes are divided into ten groups based on the importance scores.
The repetitive element families are labeled on the top.
}
\label{fig:4}
\end{figure}

Next, we sought to explore why the sequences and histone modifications identified in the previous section are important for chromatin spatial positioning.
We first inspected the genome-wide distribution of each pattern by comparing it with the SPIN states~\citep{wang2021spin}, which provides a genome-wide annotation for nuclear compartmentalization by integrating TSA-seq, DamID, and Hi-C (\textbf{Fig.}~\ref{fig:4}A), in all four cell types. 
The SPIN states include Speckle, Interior Active 1-3, Interior Repressive 1-3, Near Lamina 1-2, and Lamina.
We observed that clusters 1 and 2, with significant joint contributions from multiple histone marks, are predominantly located in the Speckle and Interior Active regions, which are found to have low LMNB TSA-seq signals \citep{wang2021spin}.
In other words, regions with these two patterns of feature contribution are generally distant from nuclear lamina, which may not be directly involved in the targeting mechanisms but instead informative in predicting regions with low TSA-seq signals.
Therefore, clusters 1 and 2 suggest that the model leverages the pattern of active histone marks such as H3K4me1 to predict the low LMNB TSA-seq regions, which is supported by the previous findings that the non lamina-associated domains (LADs) have higher enrichment of active histone marks~\citep{van2017lamina}.
Cluster 3, characterized by the high contribution of H3K9me3 and sequence features, is located exclusively in the Lamina and Near Lamina regions, consistent with the earlier findings that LADs are enriched with H3K9me3~\citep{briand2020lamina}.
Cluster 4, consisting of regions with high contribution of H3K27me3 and sequence features, is located in both interior active and repressive states. 
Recent studies~\citep{harr2015directed,chen2018mapping} revealed that H3K27me3 is highly enriched in the border of LADs and is potentially involved in LAD formation.
Thus, we hypothesize that the model relies on H3K27me3 
to predict the proximity to nuclear lamina, especially the sharp increase and decrease that indicates the formation of the LADs.
To test this hypothesis, we assessed the importance score of H3K27me3 near the LAD boundaries (\textbf{Fig.}~\ref{fig:4}B).
For all cell types, we detected a sharp peak in H3K27me3 feature importance at the LAD boundary, which is not observed for other histone marks nor in H3K27me3 for SON TSA-seq predictions.
\textbf{Fig.}~\ref{fig:4}C illustrates the clear correspondence of the H3K27me3 importance score peaks and the LAD boundaries.
Together, these analyses confirm the role of H3K27me3 in predicting the TSA-seq signals near the LAD boundaries.
Clusters 5 and 6, where only sequence features are important for prediction, are located preferentially in repressive states, including Near Lamina 1-2 and Lamina, indicating that the models rely on sequence features to predict regions lacking histone marks.

To uncover the sequence features predictive of chromosome positioning relative to nuclear lamina, we conducted a sequence composition analysis to explore what sequence elements are significantly enriched in the high-importance regions.
Particularly, we compared the sequence importance score with existing sequence annotations, including candidate cis-regulatory elements (cCRE), transcription factor (TF) binding motif, and repetitive elements.
We randomly sampled a background set of genomic regions for the purpose of comparison.
To explore whether regulatory elements are predictive of the spatial positioning of chromatin, we assessed the importance score for the top 5\% genomic regions with the highest enrichment of the cCREs annotated by SCREEN~\citep{moore2020expanded}.
We observed that regions enriched in promoters and enhancers are assigned considerably higher importance scores than background regions (\textbf{Fig.}~\ref{fig:4}D), suggesting that promoters and enhancers are potentially involved in large-scale chromatin spatial positioning.
Next, we performed TF motif enrichment analysis based on 401 motifs in the human genome collected in HOCOMOCO v11 CORE~\citep{kulakovskiy2018hocomoco}. 
We utilized FIMO~\citep{grant2011fimo} to scan through the entire genome and counted the frequency of motif ($q$-value < 0.05) within each 25kb bin.
We divided the genomic regions on the test chromosomes into ten groups based on the importance scores and calculated the motif enrichment for each group.
\textbf{Fig.}~\ref{fig:4}E shows the top motifs with the highest enrichment in the most important group for the prediction of K562 SON TSA-seq and K562 LMNB TSA-seq.
A large number of TFs belong to the Sp/KLF family.
Notably, we found that CTCF is highly enriched in the important regions, which matches the previous findings on the role of CTCF in modulating LAD boundaries~\citep{harr2015directed}. 
Additionally, we compared the sequence importance score with repetitive elements (\textbf{Fig.}~\ref{fig:4}F). 
The important regions have a higher abundance of ERVK, which is a long tandem repeat, and a moderate enrichment of the small RNAs (scRNA and srpRNA).
Consistent with previous reports on the association between LADs and LINE, LINE element L1 is slightly enriched in the most predictive loci ~\citep{meuleman2013constitutive}. 
We found that the satellite DNA is exclusively enriched in the most important regions, suggesting its highly predictive role for LMNB TSA-seq.
We repeated the same analysis for the other three cell types, and we observed that the identified important sequence elements are highly conserved across cell types.

Together, these analyses of feature importance scores and specific sequence properties further identified potential sequence and epigenomic determinants for chromatin spatial positioning to nuclear bodies.

\section*{Discussion}

In this work, we developed UNADON as a novel deep learning-based method to predict the spatial positioning of the chromosomes relative to functional nuclear bodies and explore the potential sequence and epigenomic determinants for modulating such spatial positioning in different cell types. 
UNADON takes advantage of the state-of-the-art transformer model and the multi-modal design to incorporate multi-omic data across a large context.
With the distinct neural architecture design, UNADON demonstrates strong prediction performance on individual cell types and generalizes accurately to unseen cell types.
Extensive analysis of the feature importance score reveals important regions that are consistent with previous experimental findings and further proposes potential mechanisms for the targeting of large-scale chromatin toward nuclear bodies.
Overall, UNADON establishes a novel framework for the prediction and interpretation of chromosome positioning relative to nuclear bodies, which can be extended to any large-scale sequence-based predictive modeling and analysis. 

There are a number of directions to further improve UNADON.
First, the attention mechanism inside the transformer layers can be further employed to enhance interpretability.
Attention weights have been widely adopted to probe pairwise dependencies in sequence data~\citep{vaswani2017attention}. 
However, the utility of attention as explanations is currently a subject of ongoing debate. 
It would be worthwhile to investigate whether attention mechanisms have the potential to provide novel insights into the interactions between different genomic regions in DNA sequences.
Second, the feature importance scores computed by the post-hoc methods detect only correlations, not causation.
As a result, the high-importance regions, such as clusters 1 and 2 that we identified in this work, are predictive of the LMNB TSA-seq signals but not necessarily indicative of the targeting mechanism towards the nuclear lamina.
In future work, perturbation-based methods can be leveraged to study the causal relationship between chromosome positioning and the sequence and epigenomic features.
These results can in turn enhance the predictive models.
Finally, the correlation between spatial positioning relative to different nuclear bodies and 3D genome features has been reported~\citep{chen2018mapping,belmont2022nuclear}, which makes it intriguing to build an integrative predictive model that combines spatial positioning of chromosomes and chromatin interactions.
We envision that the future versions of UNADON will connect 1D genome features, including both sequence features and epigenomic properties, with 3D chromatin structure and positioning in a more cohesive and interpretable way. 

\section*{Acknowledgements}

The authors would like to thank Yang Zhang and Andrew Belmont for their helpful discussions.
The authors are also grateful to Andrew Belmont and Steven Henikoff for making data (TSA-seq and CUT\&RUN, respectively) available before publication.
This work was supported in part by the National Institutes of Health Common Fund 4D Nucleome Program grant UM1HG011593 (J.M.) and National Institutes of Health grant R01HG007352 (J.M.).
J.M. was additionally supported by a Google Research Collabs Award and a Guggenheim Fellowship from the John Simon Guggenheim Memorial Foundation. 

\section*{Competing Interests} 

The authors declare no competing interests.

\bibliographystyle{naturemag}

\clearpage
\setcounter{page}{1}
\renewcommand\thefigure{S\arabic{figure}}
\setcounter{figure}{0}    
\renewcommand{\thetable}{S\arabic{table}}
\setcounter{table}{0}  

\begin{center}
{\Large\sffamily\bfseries 
UNADON: Transformer-based model to predict genome-wide chromosome spatial position\\
\vspace{20pt} (SUPPLEMENTAL INFORMATION)}
\end{center}

\vspace{10pt}

\appendix
\section{Supplementary table}

{\scriptsize \sf 
\begin{longtable}[c]{|ccc|}
\hline 
\multicolumn{1}{|c|}{Data}                   & \multicolumn{1}{c|}{Source}          & Identifier   \\ \hline
\endfirsthead
\multicolumn{3}{c}%
{{\bfseries Table \thetable\ continued from previous page}} \\
\hline
\multicolumn{1}{|c|}{Data}                   & \multicolumn{1}{c|}{Source}          & Identifier   \\ \hline
\endhead
\multicolumn{3}{|c|}{\textbf{K562}}                                                                \\ \hline
\multicolumn{1}{|c|}{ATAC-seq}               & \multicolumn{1}{c|}{ENCODE}          & ENCSR868FGK  \\ \hline
\multicolumn{1}{|c|}{H2A.Z ChIP-seq}         & \multicolumn{1}{c|}{ENCODE}          & ENCSR000APC  \\ \hline
\multicolumn{1}{|c|}{H3K4me1 ChIP-seq}       & \multicolumn{1}{c|}{ENCODE}          & ENCSR000EWC  \\ \hline
\multicolumn{1}{|c|}{H3K4me2 ChIP-seq}       & \multicolumn{1}{c|}{ENCODE}          & ENCSR000AKT  \\ \hline
\multicolumn{1}{|c|}{H3K4me3 ChIP-seq}       & \multicolumn{1}{c|}{ENCODE}         & ENCSR668LDD  \\ \hline
\multicolumn{1}{|c|}{H3K9me3 ChIP-seq}       & \multicolumn{1}{c|}{ENCODE}          & ENCSR000APE  \\ \hline
\multicolumn{1}{|c|}{H3K27me3 ChIP-seq}      & \multicolumn{1}{c|}{ENCODE}          & ENCSR000EWB  \\ \hline
\multicolumn{1}{|c|}{H3K27ac ChIP-seq}       & \multicolumn{1}{c|}{ENCODE}          & ENCSR000AKP  \\ \hline
\multicolumn{1}{|c|}{H3K36me3 ChIP-seq}      & \multicolumn{1}{c|}{ENCODE}          & ENCSR000AKR  \\ \hline
\multicolumn{3}{|c|}{\textbf{H1}}                                                                  \\ \hline
\multicolumn{1}{|c|}{ATAC-seq}               & \multicolumn{1}{c|}{4DN data portal} & 4DNESLMCRW2C \\ \hline
\multicolumn{1}{|c|}{H2A.Z ChIP-seq}         & \multicolumn{1}{c|}{ENCODE}          & ENCSR571IIS  \\ \hline
\multicolumn{1}{|c|}{H3K4me1 ChIP-seq}       & \multicolumn{1}{c|}{ENCODE}          & ENCSR271TFS  \\ \hline
\multicolumn{1}{|c|}{H3K4me2 ChIP-seq}       & \multicolumn{1}{c|}{ENCODE}          & ENCSR322MEI  \\ \hline
\multicolumn{1}{|c|}{H3K4me3 ChIP-seq}       & \multicolumn{1}{c|}{ENCODE}          & ENCSR443YAS  \\ \hline
\multicolumn{1}{|c|}{H3K9me3 ChIP-seq}       & \multicolumn{1}{c|}{ENCODE}          & ENCSR883AQJ  \\ \hline
\multicolumn{1}{|c|}{H3K27me3 ChIP-seq}      & \multicolumn{1}{c|}{ENCODE}          & ENCSR928HYM  \\ \hline
\multicolumn{1}{|c|}{H3K27ac ChIP-seq}       & \multicolumn{1}{c|}{ENCODE}          & ENCSR880SUY  \\ \hline
\multicolumn{1}{|c|}{H3K36me3 ChIP-seq}      & \multicolumn{1}{c|}{ENCODE}          & ENCSR476KTK  \\ \hline
\multicolumn{3}{|c|}{\textbf{HCT116}}                                                              \\ \hline
\multicolumn{1}{|c|}{ATAC-seq}               & \multicolumn{1}{c|}{ENCODE}          & ENCSR872WGW  \\ \hline
\multicolumn{1}{|c|}{H2A.Z ChIP-seq}         & \multicolumn{1}{c|}{ENCODE}          & ENCSR227XNT  \\ \hline
\multicolumn{1}{|c|}{H3K4me1 ChIP-seq}       & \multicolumn{1}{c|}{ENCODE}          & ENCSR161MXP  \\ \hline
\multicolumn{1}{|c|}{H3K4me2 ChIP-seq}       & \multicolumn{1}{c|}{ENCODE}          & ENCSR794ULT  \\ \hline
\multicolumn{1}{|c|}{H3K4me3 ChIP-seq}       & \multicolumn{1}{c|}{ENCODE}          & ENCSR333OPW  \\ \hline
\multicolumn{1}{|c|}{H3K9me3 ChIP-seq}       & \multicolumn{1}{c|}{ENCODE}          & ENCSR179BUC  \\ \hline
\multicolumn{1}{|c|}{H3K27me3 ChIP-seq}      & \multicolumn{1}{c|}{ENCODE}          & ENCSR810BDB  \\ \hline
\multicolumn{1}{|c|}{H3K27ac ChIP-seq}       & \multicolumn{1}{c|}{ENCODE}          & ENCSR661KMA  \\ \hline
\multicolumn{1}{|c|}{H3K36me3 ChIP-seq}      & \multicolumn{1}{c|}{ENCODE}          & ENCSR091QXP  \\ \hline
\multicolumn{3}{|c|}{\textbf{HFFc6}}                                                               \\ \hline
\multicolumn{1}{|c|}{ATAC-seq}               & \multicolumn{1}{c|}{4DN data portal} & 4DNESMBA9T3L \\ \hline
\multicolumn{1}{|c|}{H2A.Z CUT\&RUN}         & \multicolumn{1}{c|}{4DN data portal} & 4DNESIBPKCJK \\ \hline
\multicolumn{1}{|c|}{H3K4me1 Mint-ChIP-seq}  & \multicolumn{1}{c|}{ENCODE}          & ENCSR340XKM  \\ \hline
\multicolumn{1}{|c|}{H3K4me2 CUT\&RUN}       & \multicolumn{1}{c|}{4DN data portal} & 4DNESWK53WP1 \\ \hline
\multicolumn{1}{|c|}{H3K4me3 Mint-ChIP-seq}  & \multicolumn{1}{c|}{ENCODE}          & ENCSR639PCR  \\ \hline
\multicolumn{1}{|c|}{H3K9me3 Mint-ChIP-seq}  & \multicolumn{1}{c|}{ENCODE}          & ENCSR938NXC  \\ \hline
\multicolumn{1}{|c|}{H3K27me3 Mint-ChIP-seq} & \multicolumn{1}{c|}{ENCODE}          & ENCSR129TUY  \\ \hline
\multicolumn{1}{|c|}{H3K27ac Mint-ChIP-seq}  & \multicolumn{1}{c|}{ENCODE}          & ENCSR510VXV  \\ \hline
\multicolumn{1}{|c|}{H3K36me3 Mint-ChIP-seq} & \multicolumn{1}{c|}{ENCODE}          & ENCSR519CMW  \\ \hline
\multicolumn{3}{|c|}{\textbf{IMR-90}}                                                              \\ \hline
\multicolumn{1}{|c|}{ATAC-seq}               & \multicolumn{1}{c|}{ENCODE}          & ENCSR200OML  \\ \hline
\multicolumn{1}{|c|}{H2A.Z ChIP-seq}         & \multicolumn{1}{c|}{ENCODE}          & ENCSR124DYB  \\ \hline
\multicolumn{1}{|c|}{H3K4me1 ChIP-seq}       & \multicolumn{1}{c|}{ENCODE}          & ENCSR831JSP  \\ \hline
\multicolumn{1}{|c|}{H3K4me2 ChIP-seq}       & \multicolumn{1}{c|}{ENCODE}          & ENCSR672XZZ  \\ \hline
\multicolumn{1}{|c|}{H3K4me3 ChIP-seq}       & \multicolumn{1}{c|}{ENCODE}          & ENCSR087PFU  \\ \hline
\multicolumn{1}{|c|}{H3K9me3 ChIP-seq}       & \multicolumn{1}{c|}{ENCODE}          & ENCSR055ZZY  \\ \hline
\multicolumn{1}{|c|}{H3K27me3 ChIP-seq}      & \multicolumn{1}{c|}{ENCODE}          & ENCSR431UUY  \\ \hline
\multicolumn{1}{|c|}{H3K27ac ChIP-seq}       & \multicolumn{1}{c|}{ENCODE}          & ENCSR002YRE  \\ \hline
\multicolumn{1}{|c|}{H3K36me3 ChIP-seq}      & \multicolumn{1}{c|}{ENCODE}          & ENCSR437ORF  \\ \hline
\caption{Dataset used in this paper}
\label{table-dataset}
\end{longtable}
}
\end{document}